\documentclass[aps,prd,reprint,twocolumn,superscriptaddress,showpacs]{revtex4-1}
\usepackage{amsfonts}
\usepackage{mathrsfs}
\usepackage{amsmath}
\usepackage{color}
\usepackage{natbib}
\usepackage{graphicx}

\usepackage[scaled=2]{helvet}
\usepackage[T1]{fontenc}
\usepackage[latin9]{inputenc}
\usepackage{array}
\usepackage{textcomp}
\usepackage{multirow}
\usepackage{amsmath}
\usepackage{amssymb}
\usepackage{graphicx}
\usepackage[title]{appendix}
\usepackage{float}
\usepackage{dsfont}

\makeatletter

\providecommand{\tabularnewline}{\\}

\@ifundefined{textcolor}{}
{%
 \definecolor{BLACK}{gray}{0}
 \definecolor{WHITE}{gray}{1}
 \definecolor{RED}{rgb}{1,0,0}
 \definecolor{GREEN}{rgb}{0,1,0}
 \definecolor{BLUE}{rgb}{0,0,1}
 \definecolor{CYAN}{cmyk}{1,0,0,0}
 \definecolor{MAGENTA}{cmyk}{0,1,0,0}
 \definecolor{YELLOW}{cmyk}{0,0,1,0}
}

\usepackage[scaled=2]{helvet}\usepackage{amstext}

\makeatletter
\@ifundefined{textcolor}{}{\definecolor{BLACK}{gray}{0}\definecolor{WHITE}{gray}{1}\definecolor{RED}{rgb}{1,0,0}\definecolor{GREEN}{rgb}{0,1,0}\definecolor{BLUE}{rgb}{0,0,1}\definecolor{CYAN}{cmyk}{1,0,0,0}\definecolor{MAGENTA}{cmyk}{0,1,0,0}\definecolor{YELLOW}{cmyk}{0,0,1,0}}

\usepackage{dcolumn}
\usepackage{bm}
\usepackage{times}
\usepackage{hyperref}
\hypersetup{colorlinks=true,linkcolor=red,citecolor=blue}
\makeatother

\makeatother

\begin{document}
\title{Zigzag magnetic order in a novel tellurate compound Na$_{4-\delta}$NiTeO$_{6}$
with $\mathit{S}$ = 1 chains}
\author{Cheng Su}
\thanks{These authors contributed equally to this work.}
\affiliation{School of Physics, Beihang University, Beijing 100191, China}
\author{Xu-Tao Zeng}
\thanks{These authors contributed equally to this work.}
\affiliation{School of Physics, Beihang University, Beijing 100191, China}
\author{Yi Li}
\affiliation{Center for Advanced Quantum Studies and Department of Physics, Beijing Normal University, Beijing 100875, China}
\author{Nvsen Ma}
\affiliation{School of Physics, Beihang University, Beijing 100191, China}
\author{Zhengwang Lin}
\affiliation{School of Physics, Beihang University, Beijing 100191, China}
\author{Chuandi Zhang}
\affiliation{School of Physics, Beihang University, Beijing 100191, China}
\author{Chin-Wei Wang}
\affiliation{Australian Nuclear Science and Technology Organisation, Lucas Heights NSW 2234, Australia}
\author{Ziyu Chen}
\affiliation{School of Physics, Beihang University, Beijing 100191, China}
\author{Xingye Lu}
\affiliation{Center for Advanced Quantum Studies and Department of Physics, Beijing Normal University, Beijing 100875, China}
\author{Wei Li}
\affiliation{CAS Key Laboratory of Theoretical Physics, Institute of Theoretical Physics, Chinese Academy of Sciences, Beijing 100190, China}
\affiliation{School of Physics, Beihang University, Beijing 100191, China}
\author{Xian-Lei Sheng}
\affiliation{School of Physics, Beihang University, Beijing 100191, China}
\author{Wentao Jin}
\email{wtjin@buaa.edu.cn}

\affiliation{School of Physics, Beihang University, Beijing 100191, China}
\begin{abstract}
Na$_{4-\delta}$NiTeO$_{6}$ is a rare example in the transition-metal tellurate family of realizing an $S$ = 1 spin-chain structure. By performing neutron powder diffraction measurements, the ground-state magnetic structure of Na$_{4-\delta}$NiTeO$_{6}$ is determined. These measurements reveal that below $T\rm_{N}$ ${\sim}$ 6.8(2) K, the Ni$^{2+}$ moments form a screwed ferromagnetic (FM) spin-chain
structure running along the crystallographic $a$ axis but these FM spin chains are coupled antiferromagnetically along the $b$ and $c$ directions, giving rise to a magnetic propagation vector of $k$ = (0, 1/2, 1/2). This zigzag magnetic order is well supported by first-principles calculations. The moment size of Ni$^{2+}$ spins is determined to be 2.1(1) $\mu$$\rm_{B}$ at 3 K, suggesting a significant quenching of the orbital moment due to the crystalline electric field (CEF) effect. The previously reported metamagnetic transition near $H\rm_{C}$ ${\sim}$ 0.1 T can be understood as a field-induced spin-flip transition. The relatively easy tunability of the dimensionality of its magnetism by external parameters makes Na$_{4-\delta}$NiTeO$_{6}$ a promising candidate for further exploring various types of novel spin-chain physics.\\
\end{abstract}
\maketitle

\section{Introduction}

Since the pioneering work of Ising and Bethe \cite{Ising_25,Bethe_31}, low-dimensional magnetic systems have been attracting tremendous attention from theorists and experimentalists in the condensed
matter physics community \cite{Steiner_76,Mikeska_91,Vasiliev_18}. It is well established that in a one-dimensional (1D) spin chain composed of magnetic ions with a quantum spin $S$ = $\frac{1}{2}$ or 1, the combination of low dimensionality and strong quantum fluctuations often gives rise to various exotic quantum phenomena including spin-Peierls transitions \cite{Bray_75,Hase_93}, Bose-Einstein condensation of magnons \cite{Zapf_06,Zapf_14}, dimerized and Haldane phases \cite{Kitazawa_96,Narumi_01,Haldane_83,Sakai_90}, and fractional excitations \cite{Mourigal_13,Huang_21}. In principle, an ideal 1D spin-chain system does not order magnetically at any finite temperature because of strong quantum fluctuations, according to the well-known Mermin-Wagner theorem \cite{Mermin_66}. However, for many quasi-1D systems, a long-range magnetic order can be triggered by weak interchain exchange interactions \cite{Vasiliev_18}.

Recently, the quaternary transition-metal tellurates Ba$_{2}M$TeO$_{6}$ ($M$ = Co, Ni, and Mn) and Na$_{2}M_{2}$TeO$_{6}$  ($M$ = Co and Ni) with the magnetic $M^{2+}$ ions forming a triangular lattice and a honeycomb lattice, respectively, have emerged as popular candidates for exploring the intriguing low-dimensional magnetism. Ba$_{2}$NiTeO$_{6}$ exhibits a collinear stripe order below the antiferromagnetic (AFM) transition at $T\rm_{N}$ = 8.6 K, stablized by the magnetic frustration and large single-ion magnetic anisotropy \cite{Asai_16,Asai_17}. Similarly, Ba$_2$MnTeO$_6$ displays a collinear stripe-type AFM order below $T\rm_{N}$ $\approx$ 20 K, stablized by a nonnegligible ferromagnetic next-nearest-neighbor interlayer coupling $J_3$ \cite{Li_20}. On the other hand, the magnetism of Ba$_{2}$CoTeO$_{6}$ can be well described by the contributions from two independent subsystems including an $S$ = 1/2 triangular Heisenberg-like antiferromagnet and a frustrated honeycomb-lattice Ising-like antiferromagnet, which order at $T\rm_{N1}$ = 12 K and $T\rm_{N2}$ = 3 K, respectively \cite{Chanlert_16,Kojima_22}. The honeycomb-lattice antiferromagnets Na$_{2}$Ni$_{2}$TeO$_{6}$ and Na$_{2}$Co$_{2}$TeO$_{6}$ display more complicated magnetic behaviors \cite{Karna_17,Bera_22,Lefrancois_16,Samarakoon_21,Lin_21,Chen_21}. The magnetic structure of Ni$^{2+}$ moments ($S$ = 1) in Na$_{2}$Ni$_{2}$TeO$_{6}$ was found to follow the chirality of the alternating Na layers, showing the coexistence of a commensurate zigzag AFM order and an incommensurate AFM state below $T\rm_{N}$ = 27.5 K \cite{Bera_22}. In contrast, a recent comprehensive experimental study on single-crystal samples of Na$_{2}$Co$_{2}$TeO$_{6}$, a putative Kitaev honeycomb magnet, has revealed a new two-dimensional (2D) long-range magnetic order below 31 K, in addition to the three-dimensional (3D) order below $T\rm_{N}$ = 26.7 K that is commonly considered of zigzag nature but turns out to arise from a ''triple-$\mathbf{q}$'' ground state \cite{Chen_21}.

The compounds listed above can be classified into the geometrically frustrated magnets possessing a trigonal or hexagonal structure, and no examples of spin-chain-type magnetism were reported until the successful synthesis of the Na$_{4}M$TeO$_{6}$ ($M$ = Co, Ni and Cu) system by He $et$ $al.$ a few years ago \cite{He_17,He_18} and the recent confirmation of a canted $\uparrow\uparrow\downarrow\downarrow$ spin structure in the $S$ = 3/2 zigzag spin-chain compound BaCoTe$_{2}$O$_{7}$ by Li $et$ $al.$ using neutron diffraction \cite{Li_21}. Na$_{4}$CoTeO$_{6}$ and Na$_{4}$NiTeO$_{6}$, crystallizing in the monoclinic (space group $P\mathrm{2}\mathrm{/}c$) and triclinic (space group $P\bar{\mathrm{1}}$) symmetry, respectively, were the first systems in the transition-metal tellurate family to realize a spin-chain structure. Distorted edge-sharing CoO$_{6}$ or NiO$_{6}$ octahedra form {[}CoTeO$_{6}${]}$_{\infty}$ or {[}NiTeO$_{6}${]}$_{\infty}$ clusters surrounded by Na$^{+}$ cations running along the crystallographic $c$ or $a$ axis, thus showing uniform zigzag chains built by Co$^{2+}$ ions or alternating zigzag chains built by Ni$^{2+}$ ions, respectively \cite{He_17}. Na$_{4}$CuTeO$_{6}$, crystallizing in the monoclinic symmetry (space group $C\mathrm{2}\mathrm{/}c$), also exhibits a unique spin-chain structure built by planar CuO$_{4}$ squares and TeO$_{6}$ octahedra along the {[}101{]} direction \cite{He_18}.

\begin{figure*}[!htbp]
    \centering
\includegraphics[width=0.95\textwidth]{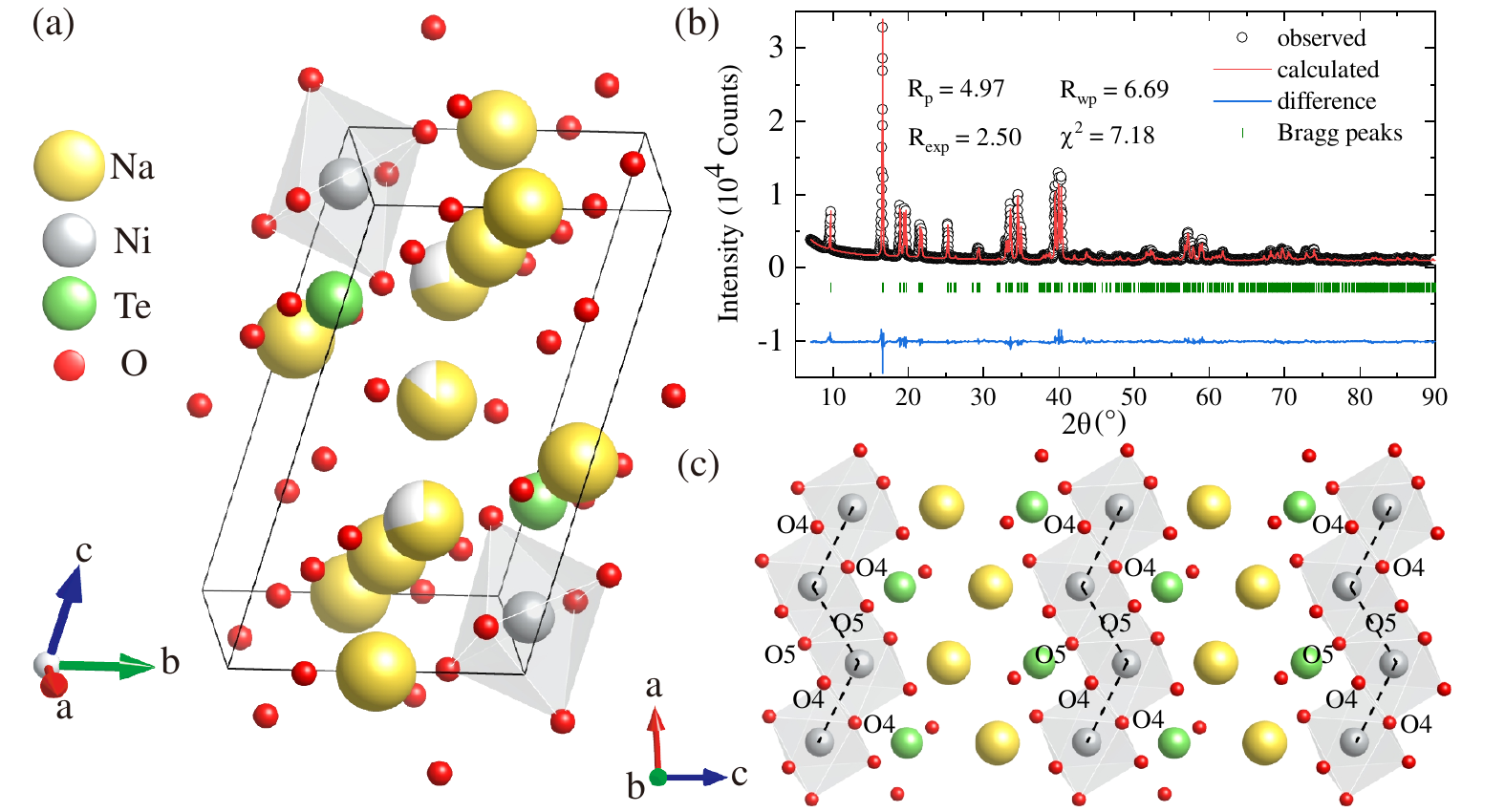}
\caption{Crystal structure of Na$_{4-\delta}$NiTeO$_{6}$ (a, c) and the XRD refinement pattern for the polycrystalline sample at room temperature (b). In (b), the black circles represent the observed intensities, and the red solid line is the calculated pattern. The difference between the observed and calculated intensities is shown as the blue line at the bottom of the figure. The expected Bragg reflections from Na$_{4-\delta}$NiTeO$_{6}$ are marked by the olive vertical bars. The dashed lines in (c) illustrate the zigzag chain structure of the Ni$^{2+}$ ions along the $a$ direction, sharing the edge of the NiO$_{6}$ octahedra in gray.}
\label{Fig1}
\end{figure*}

Despite the well-understood, distinct structures of these novel spin-chain compounds, studies on their magnetic properties are scarce, to the best of our knowledge. According to the macroscopic magnetization measurements on polycrystalline samples, the only experimental probes so far, Na$_{4}$CoTeO$_{6}$, Na$_{4}$NiTeO$_{6}$, and Na$_{4}$CuTeO$_{6}$, undergo a long-range AFM transition at $T\rm_{N}$ $\sim$ 3 K, $\sim$ 6.5 K, and $\sim$ 2.5 K, respectively \cite{He_17,He_18}. To promote the understanding of the underlying magnetic order and magnetic interactions in these novel 1D spin systems, microscopic probes of their magnetism such as neutron scattering measurements will be crucial.

In this paper, we have determined the magnetic structure of Na$_{4}$NiTeO$_{6}$, a novel tellurate compound with ${S}$ = 1 chains, using neutron powder diffraction. The Ni$^{2+}$ moments are found to form a screwed ferromagnetic (FM) spin-chain structure running along the crystallographic $a$ axis, which is coupled antiferromagnetically along the $b$ and $c$ directions, yielding a magnetic propagation vector of $k$ = (0, 1/2, 1/2). This zigzag magnetic order is supported by first-principles calculations and accounts for the dominant ferromagnetic interaction as reflected by the positive Curie-Weiss temperature. On the basis of this zigzag magnetic order, the previously reported metamagnetic transition near $H\rm_{C}$ $\sim$ 0.1 T can be understood as a field-induced spin-flip phase transition. The moment size of Ni$^{2+}$ spins is determined to be 2.1(1) $\mu$$\rm_{B}$ at 3 K, suggesting a significant quenching of the orbital moment due to the crystalline electric field (CEF) effect. The strengths of magnetic exchange couplings in this novel spin-chain compound are also discussed. 

\section{Experiments and Calculations}

Polycrystalline samples of Na$_{4}$NiTeO$_{6}$ were prepared using the traditional solid-state reaction method. The reagents of Na$_{2}$CO$_{3}$ (99.99\%), NiO (99.9\%), and TeO$_{2}$ (99.99\%) were mixed in a molar ratio of 2.4:1:1 and well ground in an agate mortar, with a slight excess of Na$_{2}$CO$_{3}$ required to compensate for the loss of sodium from evaporation. The ground powder was placed in an aluminum crucible and heated in air at 900 \textcelsius{} for 10 h. The sintering process was repeated twice to minimize the impurity phases. The attempts to use the stoichiometric ratio between the reagents led to impurity phases clearly visible in the x-ray diffraction (XRD) pattern, due to the insufficient reaction between the reagents caused by the evaporation of sodium. Using a 20\% excess of Na$_{2}$CO$_{3}$ led to a satisfactory high-purity XRD pattern.

The phase purity of the synthesized samples was checked using XRD. Powder samples with a total mass of $\sim$ 0.2 g were filled into the groove of a glass slide with a size of $\sim$ 10 $\times$10 $\times$ 0.5 mm\textsuperscript{3}, and the  XRD pattern was collected using a Bruker D8 ADVANCE in Bragg-Brentano geometry with Cu-K$_{\alpha}$ radiation ($\lambda$ = 1.5406 \AA{}) at room temperature in the range of 7-90$^{\circ}$. The specific heat and dc magnetic susceptibility of the sample were measured using a Quantum Design Physical Property Measurement System (PPMS) and Magnetic Property Measurement System (MPMS), respectively. 

Low-temperature neutron powder diffraction (NPD) measurements were performed on the high-resolution powder diffractometer Echidna at the Australian Nuclear Science and Technology Organization. Powder samples with a total mass of 6.3 g were filled into a vanadium can with a diameter of 9 mm. The diffraction patterns were recorded using incident neutrons with the wavelengths of 1.6215 and 2.4395 \AA{} at 20 K and 3 K, above and below $T\rm_{N}$, respectively. Refinements of nuclear and magnetic structures were carried out using the FULLPROF program suite \cite{Rodriguez-Carvajal_93}. As vacancies of $\sim$ 10\% were found on the Na sites according to the refinements, we will denote Na$_{4}$NiTeO$_{6}$ as Na$_{4-\delta}$NiTeO$_{6}$ below.

First-principles calculations were performed on the basis of density-functional theory (DFT) using the generalized gradient approximation (GGA) as proposed by Perdew $et$ $al$. \cite{perdew1996}, as implemented in the Vienna ab initio Simulation Package (VASP) \cite{kresse1996,kresse1999}. The energy cutoff of the plane-wave basis was set to 450 eV. The energy convergence criterion in the self-consistent calculations was set to 10$^{-5}$ eV. A $\Gamma$-centered Monkhort-Pack $k$-point mesh with a resolution of 2$\pi$ \texttimes{} 0.03 \AA{}$^{-1}$ was used for the first Brillouin zone sampling. To account for the correlation effects of Ni, we adopted the GGA + $U$ method \cite{anisimov1991} with the value of $U$ = 5 eV, which is commonly used in studying nickel compounds \cite{Bengone2000}. The DFT calculations were performed by adopting a stoichiometric composition of Na$_{4}$NiTeO$_{6}$, as the slight Na deficiency was estimated to have a minimal effect on the magnetic property (which will be discussed in detail later).

Quantum Monte Carlo (QMC) simulations were also performed to help the understanding of the finite-temperature magnetic transition (see more details in the Supplemental Material).

\section{Results and discussions}\label{sec:3}

Na$_{4-\delta}$NiTeO$_{6}$ crystallizes in a triclinic structure (space group $P\bar{\mathrm{1}}$) as reported in Ref. \cite{He_17} {[}see Fig. \ref{Fig1}(a){]}, which is confirmed by the Rietveld refinement
to the room-temperature XRD pattern of our polycrystalline sample, as shown in Fig. \ref{Fig1}(b). A possible impurity phase of Na$_{2}$CO$_{3}$, as revealed by the NPD patterns shown below, was indiscernible from the XRD results. As shown in Fig. \ref{Fig1}(c), in Na$_{4-\delta}$NiTeO$_{6}$, the magnetic Ni$^{2+}$ ions form a zigzag chain along the $a$ axis by sharing the edges (O4-O4 and O5-O5) of the NiO$_{6}$ octahedra. On the other hand, these zigzag chains of Ni$^{2+}$ ions are well separated along the $b$ or $c$ axis by the Na$^{+}$ and Te$^{6+}$ ions. For comparison, the intrachain Ni-Ni distances are determined to be 3.10757(3) and 3.08511(3) \AA{}, respectively, reflecting the nonuniform nature of the zigzag chains, while the interchain Ni-Ni distances along the $b$ and $c$ axes are much larger than 5.64272(5) and 9.52662(7) \AA{}, respectively, indicating that the dominant magnetic exchange interaction should be the intrachain coupling along $a$. 

\begin{figure}[H]
    \centering
\includegraphics[width=0.48\textwidth]{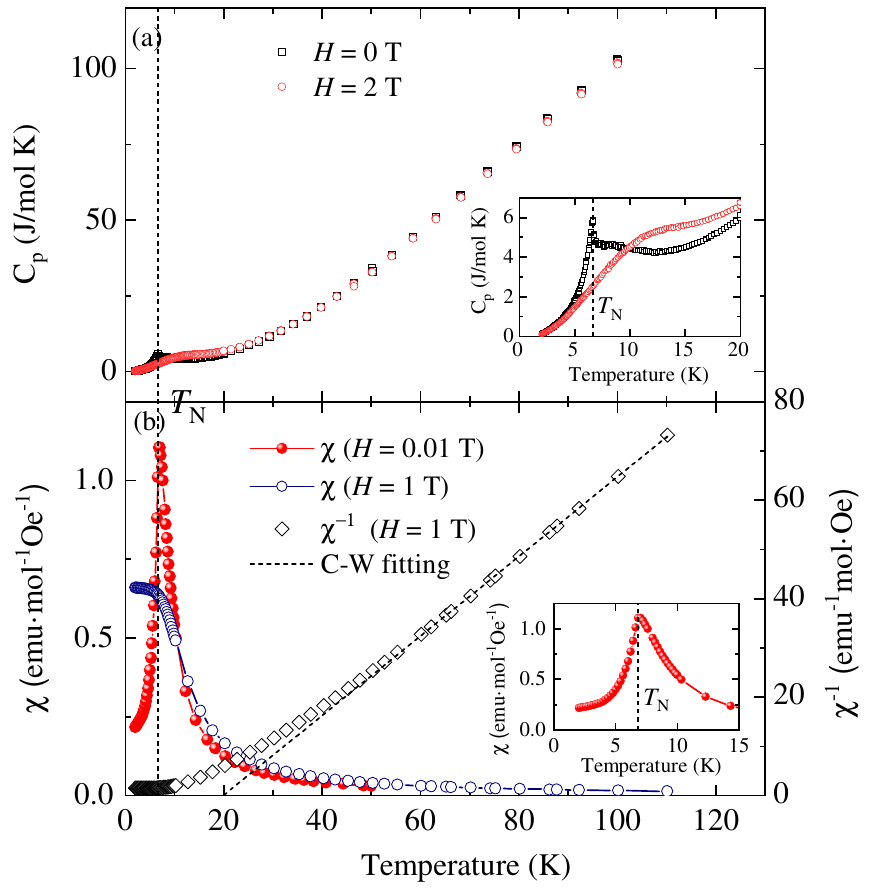}
\caption{Temperature dependencies of the molar specific heat of polycrystalline Na$_{4-\delta}$NiTeO$_{6}$ measured under zero field (black squares) and an applied magnetic field of 2 T (red circles) (a) and the dc magnetic susceptibility ($\chi$) measured in zero field cooling (ZFC) mode under an applied magnetic field of 0.01 T (red spheres) and 1 T (navy circles) (b), respectively. The insets in (a) and (b) show the enlarged region around $T\rm_{N}$. The inverse magnetic susceptibility (1/$\chi$) under $H$ = 1 T is shown by the black diamonds and the dashed line represents the Curie-Weiss fitting from 60 K to 110 K. }
\label{Fig2}
\end{figure}

Fig. \ref{Fig2}(a) shows the molar specific heat of Na$_{4-\delta}$NiTeO$_{6}$ as a function of temperature, in which a transition at 6.7(1) K can be identified in zero field, corresponding to the reported long-range AFM order \cite{He_17}. The temperature dependence of the dc magnetic susceptibility ($\chi$) of polycrystalline Na$_{4-\delta}$NiTeO$_{6}$ measured in an applied magnetic field of 0.01 T in ZFC mode is shown in Fig. \ref{Fig2}(b), clearly indicating the occurrence of an AFM transition at $T\rm_{N}$ $\sim$ 6.8(1) K, quite close to $T\rm_{N}$ $\sim$ 6.5 K reported in Ref. \cite{He_17}, below which $\chi$ decreases sharply. Combining the specific heat and dc magnetic susceptibility data, a conclusion can be reached that an AFM transition at $T\rm_{N}$ $\sim$ 6.8(2)K occurs in our polycrystalline sample. However, under $H$ = 1 T, the AFM order below $T\rm_{N}$ is suppressed, and typical FM-like behavior of the magnetic susceptibility is exhibited, consistent with that expected for FM spin chains with negligible interchain couplings. The suppression of the AFM order by the applied field is also evidenced by the specific heat measured under $H$ = 2 T, as shown in Fig. 2(a), which does not show any sharp peak.

As the dashed line in Fig. \ref{Fig2}(b) shows, a Curie-Weiss fitting to the inverse magnetic susceptibility (1/$\chi$) under $H$ = 1 T in the paramagnetic state from 60 K to 110 K yields an effective magnetic moment of $\mu\rm_{eff}$ = 3.14(1) $\mu$$\rm_{B}$ for the Ni$^{2+}$ ions and a Curie-Weiss temperature of $\theta$ = 19.8(6) K. This $\mu\rm_{eff}$ value is much closer to the spin-only moment value of $2\sqrt{S(S+1) } $ = 2.83 $\mu$$\rm_{B}$, compared with the spin-orbit coupled moment value of $g\sqrt{J(J+1)}$ = 5.59 $\mu$$\rm_{B}$, where $g$ is the Land\'e factor, indicating a significant quenching of the orbital moment, as usually expected for 3$d$ transition metal elements because of the CEF effect. The large positive value of $\theta$ suggests dominant FM interactions in this magnetic system, hinting at a possible FM coupling within the zigzag chains of the Ni$^{2+}$ ions, as the interchain interactions should be much weaker because of the much larger interchain distances along $b$ and $c$. On the basis of these results, one can speculate that the Ni$^{2+}$ moments might be ferromagnetically aligned along the $a$ direction but antiferromagnetically aligned along the $b$ and $c$ directions. Of course, this naive speculation needs the verification of microscopic probes such as magnetic neutron diffraction. 

\begin{figure*}[htb]
\centering
\includegraphics[width=0.95\textwidth]{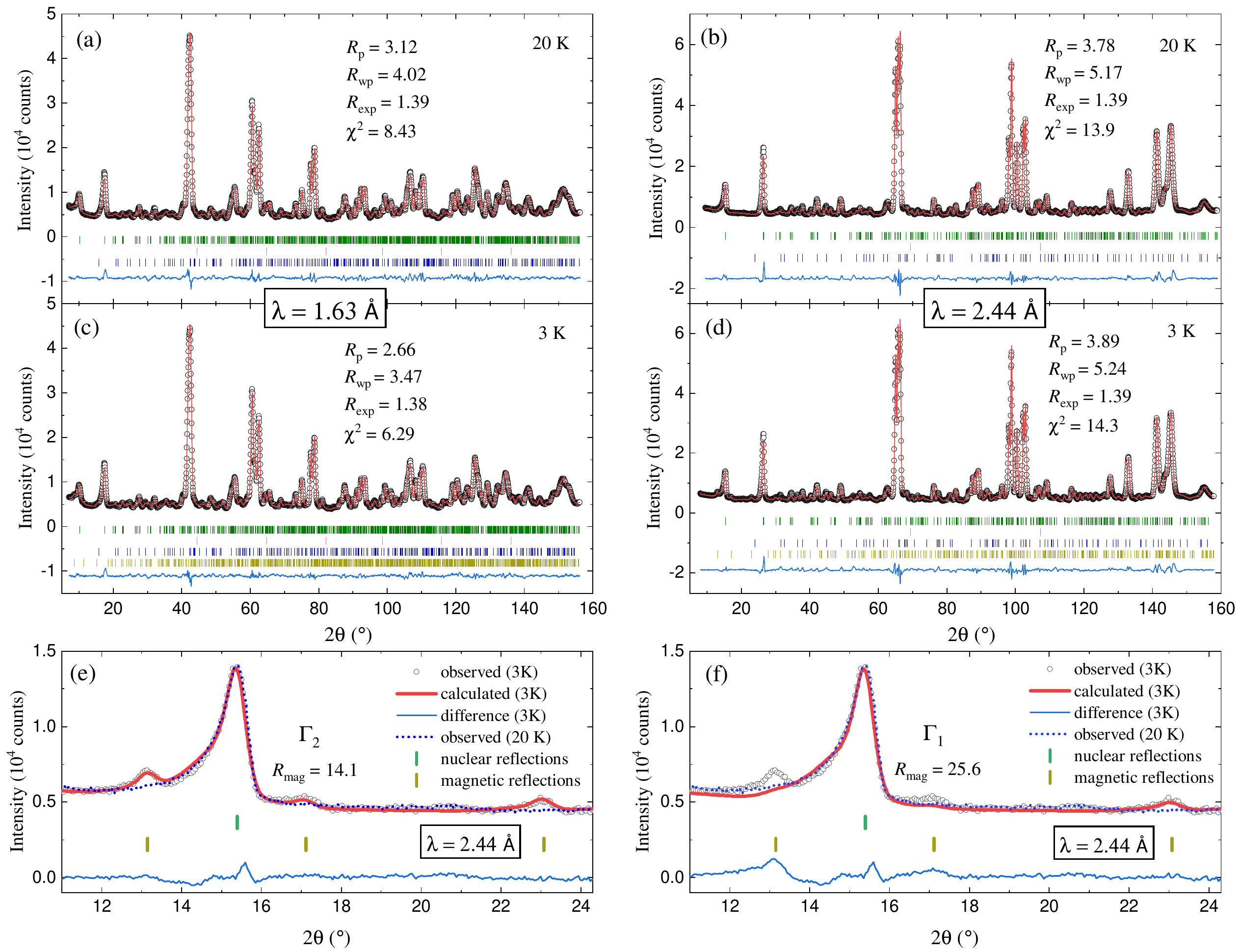}
\caption{NPD patterns of Na$_{4-\delta}$NiTeO$_{6}$ at 20 K {[}(a) and (b){]} and 3 K {[}(c) and (d){]} and the Rietveld refinements. The left {[}(a) and (c){]} and right {[}(b) and (d){]} panels show the data collected using an incident neutron wavelength of 1.63 and 2.44 \AA{}, respectively. The patterns in (c) and (d) are the refinement results obtained by adopting a magnetic structure model with the irreducible representation $\Gamma_{2}$, as described in the text. The circles represent the observed intensities, and the solid lines are the calculated patterns. The differences between the observed and calculated intensities are shown at the bottom of the figures. The vertical bars in olive, magenta, blue, and dark yellow mark the expected nuclear Bragg reflections from the Na$_{4-\delta}$NiTeO$_{6}$ main phase, vanadium sample container, Na$_{2}$CO$_{3}$ impurity, and the magnetic Bragg reflections from Na$_{4-\delta}$NiTeO$_{6}$, respectively. (e) and (f) show the enlarged high-resolution ($\lambda$ = 2.44 \AA{}) diffraction patterns at 3 K, fitted using the magnetic structure described by $\Gamma_{2}$ and $\Gamma_{1}$, respectively, in which the observed intensities at 3 K and 20 K and the calculated intensities at 3 K are shown for comparison.}
\label{Fig3}
\end{figure*}

\begin{table}[H]
\centering
\caption{Refinement results of the structural parameters of \textcolor{black}{Na$_{4-\delta}$NiTeO$_{6}$ determined by NPD at 20 K, including the atomic coordinates and site occupancies.}}
\label{table1}

\resizebox{0.45\textwidth}{!}{
\begin{tabular}{cccccc}
    \hline\hline
Atom/site & x & y & z & $B_{iso}$ ($\AA{}^{2}$) & occupancy\tabularnewline
\hline 
Te (2$i$) & 0.243(1) & 0.992(2) & 0.2456(9) & 0.1(1) & 1.01(1)\tabularnewline
Ni (2$i$) & 0.249(1) & 1.005(1) & -0.0809(1) & 0.46(6) & 1.01(1)\tabularnewline
O1 (2$i$) & -0.085(1) & 1.194(2) & 0.2014(8) & 0.23(3) & 0.99(1)\tabularnewline
O2 (2$i$) & 0.342(2) & 1.212(2) & 0.3427(8) & 0.23(3) & 1.00(1)\tabularnewline
O3 (2$i$) & 0.587(2) & 0.792(2) & 0.2655(8) & 0.23(3) & 0.98(1)\tabularnewline
O4 (2$i$) & 0.372(2) & 1.198(2) & 0.0544(8) & 0.23(3) & 1.03(1)\tabularnewline
O5 (2$i$) & 0.120(2) & 0.802(2) & 0.1200(7) & 0.23(3) & 0.99(1)\tabularnewline
O6 (2$i$) & 0.149(2) & 0.770(2) & 0.4208(7) & 0.23(3) & 0.98(1)\tabularnewline
Na1 (1$c$) & 0 & 0.5 & 0 & 0.21(4) & 0.52(2)\tabularnewline
Na2 (2$i$) & 0.501(2) & 0.496(3) & 0.168(1) & 0.21(4) & 1.04(3)\tabularnewline
Na3 (2$i$) & 0.982(3) & 0.502(4) & 0.322(2) & 0.21(4) & 0.71(3)\tabularnewline
Na4 (1$h$) & 0.5 & 0.5 & 0.5 & 0.21(4) & 0.43(2)\tabularnewline
Na5 (2$i$) & 0.242(3) & 0.004(4) & 0.584(2) & 0.21(4) & 1.01(2) \tabularnewline
\hline\hline
\end{tabular}
}
\end{table}

To verify our speculation about the ground-state spin configuration in Na$_{4-\delta}$NiTeO$_{6}$, NPD experiments were performed for the magnetic structure determination. Two sets of neutron diffraction
patterns were collected, with the wavelength of 1.62 $\AA{}$ providing a wider $Q$-coverage suitable for refining the thermal factors and site occupancies and the wavelength of 2.44 $\AA{}$ enabling a higher resolution for precisely determining the magnetic propagation vector $k$. Fig. \ref{Fig3} summarizes the NPD data of Na$_{4-\delta}$NiTeO$_{6}$ collected using these two wavelengths at 20 K (in the paramagnetic state, a and b) and 3 K (in the AFM state, c and d), respectively, as well as the refinement results for comparison. By a simultaneous refinement of the NPD patterns recorded at 20 K using two wavelengths, the structural parameters including the atomic coordinates and site occupancies can be refined, as listed in Table \ref{table1}. Noticeably, the refined occupancies of Na3 and Na4 sites are evidently lower than the stoichiometric case of 1 and 0.5, respectively, which cannot be resolved by XRD because of the low x-ray scattering cross section of Na atoms. However, fixing the occupancies of all Na atoms to the stoichiometric values can also well fit the patterns with very tiny increases in the $R$ factors ($R_{\mathrm{p}}$ = 3.25 and $R_{\mathrm{wp}}$ = 4.19 for $\lambda$ = 1.62 \AA{}, and $R_{\mathrm{p}}$ = 3.98 and $R_{\mathrm{wp}}$ = 5.41 for $\lambda$ = 2.44 \AA{}), suggesting that the real level of Na deficiency cannot be precisely determined with our current data collected using thermal neutrons because of insufficient coverage of high-$Q$ reflections. On the basis of the refined occupancies of different Na sites, the nominal composition of our sample is denoted as \textcolor{black}{Na$_{3.7}$NiTeO$_{6}$}. In addition to this main phase, multiple unexpected peaks indicative of the Na$_{2}$CO$_{3}$ impurity phase ($\sim$ 8\% wt) were also observed. Because the same sample was used in the XRD and NPD measurements and almost no trace of Na$_{2}$CO$_{3}$ can be observed in the XRD pattern {[}see Fig. \ref{Fig1}(b){]}, we speculate that this discrepancy is due to the significantly different penetration depths of x-rays and neutrons into the sample or the slight inhomogeneity of the sample.

At 3 K, which is below $T\rm_{N}$ $\sim$ 6.8 K, additional magnetic reflections due to the AFM ordering of Ni$^{2+}$ moments were observed, as clearly shown in Figs. \ref{Fig3}(e) and \ref{Fig3}(f). Using the
K\_SEARCH program integrated into the FULLPROF suite, the magnetic propagation vector of $k$ = (0, 1/2, 1/2) is determined, and all strong magnetic reflections in the high-resolution diffraction pattern collected with an incident neutron wavelength of 2.44 \AA{} can be well indexed. According to the representation analysis performed for the space group of $P\bar{1}$ using the BASIREPS program also integrated into the FULLPROF suite, the magnetic representation $\Gamma\rm_{mag}$ for the Ni (2$i$) site can be decomposed as the sum of two irreducible representations (IRs), 
$$\Gamma_{\mathrm{mag}}=3\Gamma_{1}^{1}\oplus3\Gamma_{2}^{1},$$ 
whose basis vectors are listed in Table \ref{table2}. 

\begin{table}[H]
    \centering
\caption{Basis vectors of the IRs for the Ni atoms in Na$_{4-\delta}$NiTeO$_{6}$ with the space group $P\bar{1}$ and $k$ = (0, 1/2, 1/2) obtained from representation analysis. The Ni atoms in one chemical unit cell are defined as Ni(1) and Ni(2), located at ($x$, $y$, $z$) and (1$-x$, 1$-y$, 1$-z$), respectively. The magnetic R values ($R\rm_{mag}$) of the fitting for each IR are also listed.}
\label{table2}
\begin{tabular}{cccccc}
    \hline\hline
IRs & $R\rm_{mag}$ & $\psi_{\nu}$ & Components & Ni(1) & Ni(2)\tabularnewline
\hline 
$\Gamma_{1}^{1}$ & 25.6 & $\psi_{1}$ & Real & (1,0,0) & (-1,0,0)\tabularnewline
 &  & $\psi_{2}$ & Real & (0,1,0) & (0,-1,0)\tabularnewline
 &  & $\psi_{3}$ & Real & (0,0,1) & (0,0,-1)\tabularnewline
\hline 
$\Gamma_{2}^{1}$ & 14.1 & $\psi_{1}$ & Real & (1,0,0) & (1,0,0)\tabularnewline
 &  & $\psi_{2}$ & Real & (0,1,0) & (0,1,0)\tabularnewline
 &  & $\psi_{3}$ & Real & (0,0,1) & (0,0,1)\tabularnewline
 \hline\hline
\end{tabular}
\end{table}
\noindent 

\begin{figure*}[htb]
\centering
\includegraphics[width=0.95\textwidth]{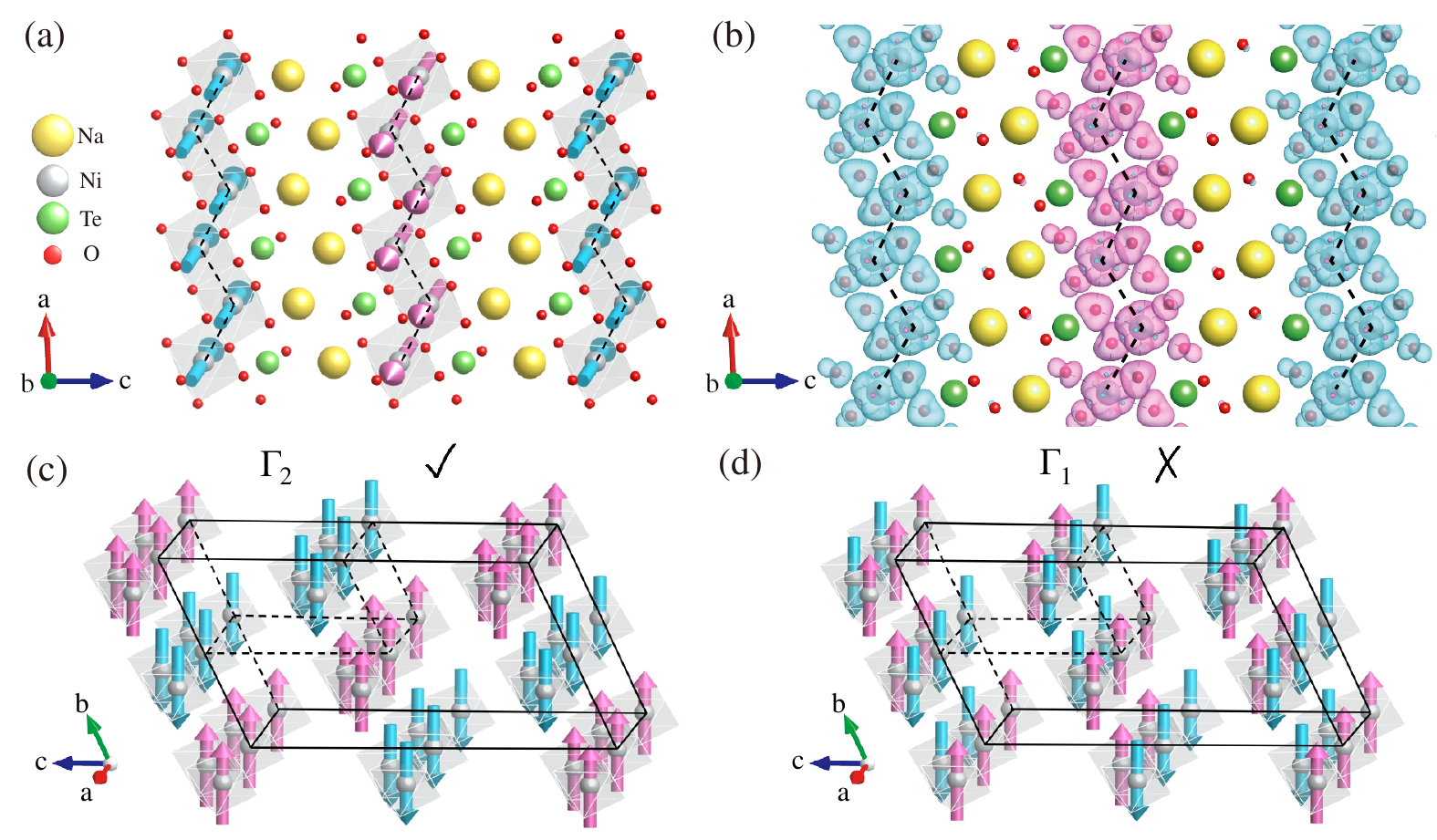}
\caption{Experimentally determined magnetic structure of Na$_{4-\delta}$NiTeO$_{6}$ viewed along the $b$ axis (a) and $a$ axis (c) described by the irreducible representation $\Gamma_{2}$. In (a), the blue
and red arrows correspond to the moments pointing in opposite directions and the dashed lines mark the zigzag chains along the $a$ axis. The two possible magnetic structures of Ni described by $\Gamma_{2}$
and $\Gamma_{1}$ are depicted in (c) and (d), respectively, allowing an FM or an AFM intrachain coupling between the nearest-neighbor Ni$^{2+}$ spins along $a$. The chemical and magnetic unit cells are
marked out by the inner dashed lines and outer solid lines in (c) and (d), respectively. The calculated spin density distribution is shown in (b) for comparison with (a), where the blue and red colors around the Ni$^{2+}$ ions represent the densities of spin-up and spin-down electrons, respectively.}
\label{Fig4}
\end{figure*}

The two distinct magnetic structures described by $\Gamma_{1}$ and $\Gamma_{2}$ are illustrated in Fig. \ref{Fig4}(a, c) and \ref{Fig4}(d), respectively. $\Gamma_{1}$ allows the two Ni$^{2+}$ ions within a primitive unit cell located at ($x$, $y$, $z$) and ($1-x$, $1-y$, $1-z$), respectively, to have antiparallelly aligned magnetic moments, corresponding to intrachain AFM couplings and interchain AFM couplings. In contrast, $\Gamma_{2}$ allows the moments of two Ni$^{2+}$ ions at ($x$, $y$, $z$) and (1$-x$, 1$-y$, 1$-z$) to align parallelly, corresponding to intrachain FM couplings and interchain AFM couplings. Both models were tested for the refinements of magnetic reflections. As shown in Figs. \ref{Fig3}(e) and \ref{Fig3}(f), the magnetic structure model given by $\Gamma_{2}$ yields an obviously better agreement between the calculated and experimental intensities of the magnetic peaks, with a lower \textcolor{black}{magnetic $R$ factor ($R\rm_{mag}$)}. The zigzag magnetic \noindent order described by $\Gamma_{2}$, in which the Ni$^{2+}$ moments show FM intrachain couplings but AFM interchain couplings, is further supported by the spin density distribution given by the DFT calculation, as shown in Fig. \ref{Fig4}(b). The density of spin-up or spin-down electrons around the Ni$^{2+}$ ions alternates along the $b$ and $c$ directions, but this density alternation does not occur along $a$, consistent with the zigzag-type magnetic order determined experimentally.

By fixing the nuclear structural parameters and the scale factor derived from the refinement of 20 K data, the coefficients of the three basis vectors ($\psi_{\nu}$, see Table \ref{table2}) of the magnetic representation $\Gamma_{2}$ were determined by refining the NPD patterns at 3 K. The lattice constants are found to remain almost invariant between 20 K and 3 K, as shown in Table \ref{table3}. The moment size of the Ni$^{2+}$ ions at 3 K along the $a$, $b$, and $c$ directions converges to 0.9(1), $-$2.14(9), and 0.65(7) $\mu$$\rm_{B}$, respectively, corresponding to a total moment of $\mu_{\mathrm{total}}$ = 2.1(1) $\mu$$\rm_{B}$. This moment value is well consistent with that expected for a spin-only $S$ = 1 Ni$^{2+}$ moment showing an ordered moment of $gS$ = 2 $\mu$$\rm_{B}$, adopting the Land\'e factor $g$ of 2 for spin-only cases. Therefore, it further corroborates the significant quenching of the orbital moment for the Ni$^{2+}$ ions, as evidenced by the magnetic susceptibility measurements. \textcolor{black}{This quenching behavior of the }Ni$^{2+}$ orbital moment in NiO$_{6}$ is well documented for various nickel compounds, including Sr$_{2}$NiMoO$_{6}$, Ba$_{2}$NiMoO$_{6}$, Li$_{2}$Ni(SO$_{4}$)$_{2}$, and Li$_{2}$Ni(WO$_{4}$)$_{2}$, showing a spin-only ordered moment in between 1.8 and 2.2 $\mu$$\rm_{B}$, as also determined by neutron diffraction \cite{Martinez-Lope_03,Reynaud_14,Ranjith_16}.

\begin{table}[H]
    \centering
\caption{\textcolor{black}{Refined lattice constants of Na$_{4-\delta}$NiTeO$_{6}$ at 20 K and 3 K and the moment size of }the Ni$^{2+}$ ions at 3 K along the $a$ ($\mu_{a}$), $b$ ($\mu_{b}$), and $c$ ($\mu_{c}$) directions. }
\label{table3}
\begin{tabular}{ccc}
    \hline\hline
 & \textcolor{black}{20 K} & 3 K\tabularnewline
\hline 
$a$ ($\textrm{\AA{}}$) & 5.38768(5) & 5.38697(4)\tabularnewline
$b$ ($\textrm{\AA{}}$) & 5.64335(5) & 5.64272(5)\tabularnewline
$c$ ($\textrm{\AA{}}$) & 9.52730(8) & 9.52662(7)\tabularnewline
$\alpha$ ($^{\circ}$) & 73.1118(9) & 73.1146(9)\tabularnewline
$\beta$ ($^{\circ}$) & 89.7006(9) & 89.6981(8)\tabularnewline
$\gamma$ ($^{\circ}$) & 80.554(1) & 80.5579(9)\tabularnewline
\hline 
$\mu_{a}$ ($\mu$$\rm_{B}$) & - & 0.9(1)\tabularnewline
$\mu_{b}$ ($\mu$$\rm_{B}$) & - & -2.14(9)\tabularnewline
$\mu_{c}$ ($\mu$$\rm_{B}$) & - & 0.65(7)\tabularnewline
$\mu\rm_{total}$ ($\mu$$\rm_{B}$) & - & 2.1(1)\tabularnewline
\hline\hline
\end{tabular}
\end{table}

As shown in Fig.~\ref{Fig4}(a) and Fig.~\ref{Fig4}(c), Na$_{4-\delta}$NiTeO$_{6}$ displays an easy-axis-type ground-state magnetic structure. The Ni$^{2+}$ moments lie mostly along the $b$ axis of the triclinic unit cell, with a deviation of {$\sim$ 30$^{\circ}$. The magnetic easy-axis can also be confirmed by calculating the magnetic anisotropy energy (MAE) of a primitive unit cell of Ni$^{2+}$ moments}.

\begin{figure*}[htp]
\centering
\includegraphics[width=0.95\textwidth]{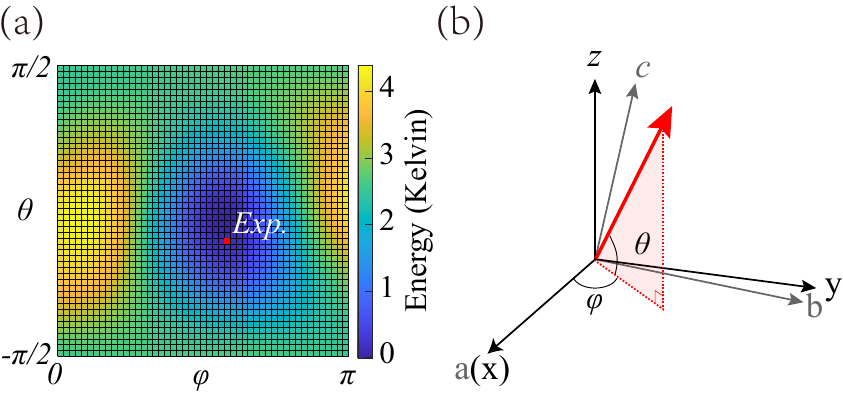}
\caption{Angular dependence of the MAE (a), with the definitions of $\theta$ and $\phi$ illustrated in (b). In (a), the color represents the relative energy of certain spin orientation calculated by DFT with respect to the magnetic ground state. The orientation of the moment direction found experimentally is labeled by the red square, which is very close to the center of the minimum region of the MAE. In (b), $a$, $b$, and $c$ mark the crystallographic axes of the triclinic unit cell, while $x$, $y$, and $z$ are the axes of the Cartesian coordinate system, with $x$ overlapping the $a$ axis. $\theta$ and $\phi$ are the angles of the spherical polar coordinate system, as illustrated. }
\label{Fig5}
\end{figure*}

Using DFT, we have calculated the energies of 801 different spin orientations uniformly distributed in the real space. Fig.~\ref{Fig5} shows the angular dependence of the MAE obtained by linear interpolations to the calculated values, with respect to the magnetic ground state.  The orientation with the moment components of (0.5283, -2.0062, 0.3255) $\mu$$\rm_{B}$ in Cartesian coordinates is found to have the lowest energy, with a small angle of $\sim$ 25.7$^{\circ}$ deviating from the $b$ axis. This calculated magnetic easy-axis direction is perfectly consistent with the experimentally determined moment orientation, with a tiny difference of $\sim$ 8.5$^{\circ}$. Note that the experimentally determined moment direction is very close to the center of the minimum region of the MAE, as shown in Fig.~\ref{Fig5}(a). Considering the negligible spin-orbit coupling due to the significant quenching of the Ni$^{2+}$ orbital moment, this easy-axis-type magnetic anisotropy in Na$_{4-\delta}$NiTeO$_{6}$ most likely arises from the single-ion anisotropy associated with the $S$ = 1 Ni$^{2+}$ ions, as documented in various nickel compounds \cite{PhysRevLett.72.3108,Zvyagin_07,Brambleby_17,Curley_21}.

\begin{figure*}[htp]
\centering
\includegraphics[width=0.95\textwidth]{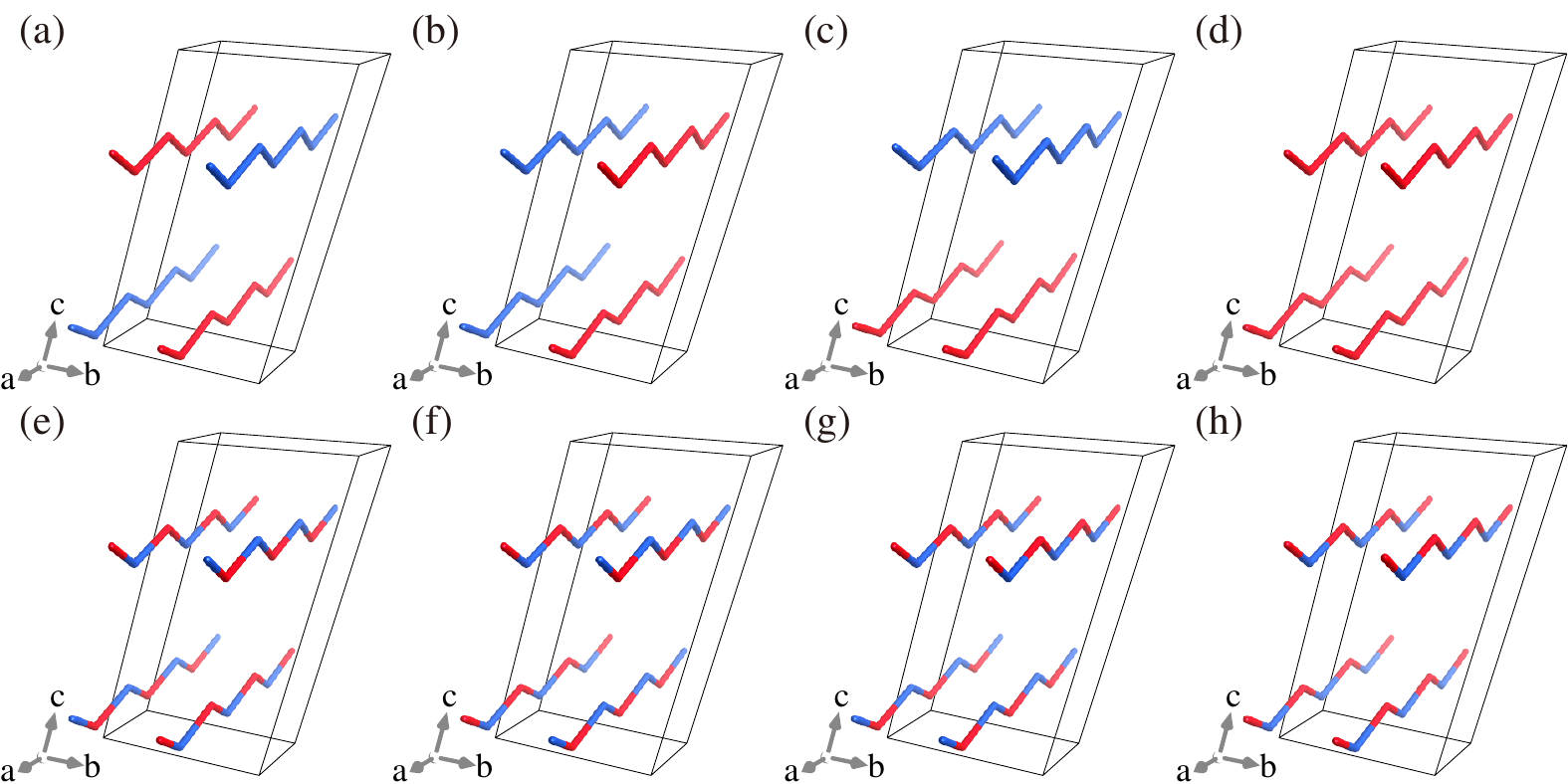}
\caption{Eight magnetic configurations of Na$_{4-\delta}$NiTeO$_{6}$, as discussed in Table \ref{table4}, including (a) G-type AFM (with FM intrachain coupling), (b) C-type AFM (FM intrachain), (c) A-type AFM (FM intrachain), (d) FM interchain (FM intrachain), (e) G-type AFM (AFM intrachain), (f) C-type AFM (AFM intrachain), (g) A-type AFM (AFM intrachain) and (h) FM interchain (AFM intrachain). Red and blue indicate the opposite moment directions of Ni$^{2+}$ ions. }
\label{Fig6}
\end{figure*}

\begin{table*}[t]
    \centering
\caption{Relative energies (in units of meV) of eight magnetic configurations for Na$_{4-\delta}$NiTeO$_{6}$ calculated using GGA+SOC+$U$. The converged magnetic moments (in units of $\mu$$\rm_{B}$) of Ni$^{2+}$ ions are also provided.}
\label{table4}
\resizebox{\textwidth}{!}{
\begin{tabular}{ccccc}
    \hline\hline
Magnetic configuration & Total moment ($\mu\rm_{B}$) & Moment components & Energy (meV) & Configuration in Fig. \ref{Fig6}\tabularnewline
\hline 
G-type AFM (FM intrachain) & 1.7210 & (${\pm}$0.7380, ${\mp}$1.7607, ${\pm}$0.5377) & 0 & a\tabularnewline
C-type AFM (FM intrachain) & 1.7210 & (${\pm}$0.7370, ${\mp}$1.7607, ${\pm}$0.5377) & 0.25 & b\tabularnewline
A-type AFM (FM intrachain) & 1.7210 & (${\pm}$0.7380, ${\mp}$1.7604, ${\pm}$0.5367) & 0.46 & c\tabularnewline
FM interchain (FM intrachain) & 1.7210 & (${\pm}$0.7370, ${\mp}$1.7604, ${\pm}$0.5367) & 0.76 & d\tabularnewline
G-type AFM (AFM intrachain) & 1.7249 & (${\pm}$0.7397, ${\mp}$1.7651, ${\pm}$0.5387) & 36.64 & e\tabularnewline
C-type AFM (AFM intrachain) & 1.7249 & (${\pm}$0.7397, ${\mp}$1.7651, ${\pm}$0.5387) & 36.83 & f\tabularnewline
A-type AFM (AFM intrachain) & 1.7249 & (${\pm}$0.7397, ${\mp}$1.7648, ${\pm}$0.5377) & 36.06 & g\tabularnewline
FM interchain (AFM intrachain) & 1.7249 & (${\pm}$0.7396, ${\mp}$1.7641, ${\pm}$0.5387) & 36.2 & h\tabularnewline
\hline\hline
\end{tabular}
}
\end{table*}

Along the $a$ axis, the Ni$^{2+}$ spins form a screwed zigzag chain with intrachain FM couplings, which explains the positive Curie-Weiss temperature $\theta$, as shown in Fig. \ref{Fig2}(b). These zigzag
chains are further antiparallelly coupled along the $b$ and $c$ axes, giving rise to an overall AFM ground state. The stability of this zigzag-type magnetic order is further verified by first-principles
calculations. To facilitate the study of magnetic configurations from the perspective of DFT, here we compare the free energies of Na$_{4-\delta}$NiTeO$_{6}$ with eight magnetic configurations. Figs.~\ref{Fig6}(a)--(h) illustrate the framework of Ni atoms forming four zigzag chains with eight spin configurations. Our calculations shown in Table \ref{table4} confirm that the lowest energy
is achieved by the G-type AFM configuration with FM intrachain couplings, well consistent with our experimental result. In terms of energy, Table \ref{table4} shows that the FM intrachain couplings are
more stable with respect to the AFM interchain couplings, suggesting the dominant interaction in Na$_{4-\delta}$NiTeO$_{6}$ is the intrachain ferromagnetism. The converged total moment of Ni$^{2+}$ ions is 1.721 $\mu$$\rm_{B}$, ${\sim}$ 18\% smaller than the experimental value of 2.1(1) $\mu$$\rm_{B}$ determined from NPD. However, it is well recognized that DFT usually underestimates the moment size for 3d transition metals by transferring some spin densities onto the surrounding O$^{2-}$ ions.

To clarify if Ni$^{3+}$ ions appear as a response to the deficiency of Na, we have performed bond valence sum (BVS) calculations using the Bond\_Str program in the FULLPROF suite based on the structural parameters listed in Table \ref{table1} and reported in Ref. \cite{He_17}, which yield a BVS value of +1.78 and +1.88, respectively, for the Ni ions. Although the BVS calculation provides only a rough estimate of the ionic valence, this result somehow reflects that the Na deficiency in our sample is small. In addition, note that the Ni atoms are well protected by the surrounding O atoms from the charge compensation effect, which may further reduce the influence of possible Na vacancies on the valence of Ni ions. In fact, the ordered moment of the Ni ions with a size of $\sim$ 2.1(1) $\mu$$\rm_{B}$ at 3 K determined by our magnetic structure refinement and an effective magnetic moment of $\sim$ 3.14(1) $\mu$$\rm_{B}$ obtained by the Curie-Weiss fitting to the inverse magnetic susceptibility in the paramagnetic range are well consistent with the values expected for Ni$^{2+}$ ions. For these reasons, we believe that the appearance of Ni$^{3+}$ ions due to the slight deficiency of Na, if any, would have a minimal effect.

According to the experimentally determined magnetic structure and the calculation results shown in Fig.~\ref{Fig5}, the magnetic easy axis is close to the $b$ axis, while the $a$ axis with the large MAE energy corresponds to the magnetic hard axis. This result implies that the ground-state spin Hamiltonian of Na$_{4}$NiTeO$_{6}$ might be approximately described by a 1D Heisenberg model $H=-J\sum_{\langle i,j\rangle}\boldsymbol{S}_{i}\cdot\boldsymbol{S}_{j}$, where the positive magnetic exchange constant $J$ for nearest-neighbor Ni$^{2+}$ spins tends to facilitate an FM coupling within the zigzag chains. The relative energies in Table \ref{table4} helped us estimate the exchange coupling constants (see the Supplemental Material for more details), yielding the intrachain couplings $J_{1}$ $\sim$ 25.1 K and $J_{1}'$ $\sim$ 24.0 K, corresponding to the two nonequivalent intrachain Ni-Ni distances, as well as interchain couplings $J_{2}$ $\sim -$0.33 K and $J_{3}$ $\sim$ $-$0.18 K, along $b$ and $c$, respectively. The estimate of $J_{1}$ is well consistent with the positive Curie-Weiss temperature of $\theta$ = 19.8(6) K. At first glance, such weak interchain interactions estimated by DFT hardly seem to account for the observed long-range AFM ordering at a relatively high temperature of $T\rm_{N}$ $\sim$ 6.8 K. To answer this interesting question, we performed QMC simulations adopting $J_{2}$ = 0.01 $J_{1}$, suggesting that the magnetic order can be realized at a temperature much higher than  $J_{2}$ in such a system with strong interchain correlations (detailed discussions are included in the Supplemental Materials). Further inelastic neutron scattering experiments will be crucial for precisely determining of the underlying spin Hamiltonians.

In addition, we note that the isothermal magnetization of this compound at 2 K shown in Ref. \cite{He_17} suggests a metamagnetic transition at $H\rm_{C}$ $\sim$ 0.1 T. This metamagnetic transition can now be understood to result from spin flipping, as the weak interchain AFM couplings should be suppressed by a small external magnetic field, and a transition from a 3D antiferromagnet to a quasi-1D FM spin-chain compound will be induced. This scenario is well supported by the distinct behaviors of our dc magnetic susceptibility data measured at $H$ = 0.01 T and $H$ = 1 T, below and above $H\rm_{C}$, respectively, as shown in Fig. \ref{Fig2}(b). The relatively easy tunability of the dimensionality of its magnetism by an external field makes this compound a promising candidate for further exploring of various types of novel spin-chain physics.\\

\section{Conclusions}\label{sec:4}
In summary, systematic studies on the magnetic properties of a novel tellurate compound Na$_{4-\delta}$NiTeO$_{6}$ with $S$ = 1 chains were performed using magnetic susceptibility measurements, neutron diffraction experiments, first-principles calculations, and QMC simulations. In the ground state, the Ni$^{2+}$ moments were found to form a screwed FM zigzag spin-chain structure running along the crystallographic $a$ axis, but these FM spin chains are coupled antiferromagnetically along the $b$ and $c$ directions, giving rise to a magnetic propagation vector of $k$ = (0, 1/2, 1/2), which is well supported by first-principles calculations. The previously reported metamagnetic transition near $H\rm_{C}$ $\sim$ 0.1 T can now be understood as a field-induced spin-flip transition. The spin Hamiltonian of Na$_{4-\delta}$NiTeO$_{6}$ is expected to be dominated by a Heisenberg-type FM coupling within the zigzag chains, while the weak interchain antiferromagnetic couplings are also important for stabilizing the antiferromagnetic zigzag order. Further inelastic neutron scattering experiments on Na$_{4-\delta}$NiTeO$_{6}$ will be useful for mapping out the spin-wave dispersion for a precise determination of the magnetic exchange couplings, which is crucial for completely understanding the magnetism in this novel tellurate compound with $S$ = 1 chains.

\begin{acknowledgments}
This work is based on experiments performed at the Australian Nuclear Science and Technology Organization (ANSTO), Sydney, Australia. The authors acknowledge Shoushu Gong and Kan Zhao for valuable discussions, Zirong Ye for the technical assistance in the MPMS measurements, and the computational support from HPC of the Beihang University. This work is financially supported by the National Natural Science Foundation of China (Grant No. 12074023, No. 12004020, No. 11834014, No. 11974036, No. 12074024) and the Fundamental Research Funds for the Central Universities in China. \\

\end{acknowledgments}

\bibliographystyle{apsrev4-1}
\bibliography{NNTO}

\end{document}